\documentclass[journal]{vgtc}                %

\ifpdf%
  \pdfoutput=1\relax                   %
  \usepackage{graphicx}                %
  \DeclareGraphicsExtensions{.pdf,.png,.jpg,.jpeg} %
\else%
  \usepackage{graphicx}                %
  \DeclareGraphicsExtensions{.eps}     %
\fi%

\graphicspath{{figures/}{pictures/}{images/}{./}} %

\usepackage{microtype}                 %
\PassOptionsToPackage{warn}{textcomp}  %
\usepackage{textcomp}                  %
\usepackage{mathptmx}                  %
\usepackage{times}                     %
\usepackage{cite}                      %
\usepackage{tabu}                      %
\usepackage{booktabs}                  %
\usepackage{subcaption}
\usepackage{multirow}
\usepackage{array}
\usepackage{colortbl}
\usepackage{xcolor}
\usepackage{tikz}
\usepackage{bm}
\usepackage{amssymb}

\newcommand\hl[1]{%
  \bgroup
  \hskip0pt\color{black}%
  #1%
  \egroup
}%

\onlineid{0}

\vgtccategory{Research}
\vgtcpapertype{please specify}

\title{Improving Visualization Interpretation Using Counterfactuals}

\author{Smiti Kaul, David Borland, Nan Cao, David Gotz}
\authorfooter{
\item
 Smiti Kaul is with the Dept. of Computer Science at the University of North Carolina at Chapel Hill. E-mail: smiti@live.unc.edu.
\item
 David Borland is with RENCI at the University of North Carolina at Chapel Hill. E-mail: borland@renci.org.
\item
 Nan Cao is with the Intelligent Big Data Visualization Lab at Tongji University. E-mail: nan.cao@tongji.edu.cn
\item
  David Gotz is with the School of Information and Library Science at the University of North Carolina at Chapel Hill. E-mail: gotz@unc.edu.
}

\shortauthortitle{Kaul \MakeLowercase{\textit{et al.}}: CoFact: Improving Visualization Interpretation Using Counterfactuals}

\abstract{
Complex, high-dimensional data is used in a wide range of domains to explore problems and make decisions. Analysis of high-dimensional data, however, is vulnerable to the hidden influence of confounding variables, especially as users apply ad hoc filtering operations to visualize only specific subsets of an entire dataset. Thus, visual data-driven analysis can mislead users and encourage mistaken assumptions about causality or the strength of relationships between features. This work introduces a novel visual approach designed to reveal the presence of confounding variables via counterfactual possibilities during visual data analysis. It is implemented in CoFact, an interactive visualization prototype that determines and visualizes \textit{counterfactual subsets} to better support user exploration of feature relationships. Using publicly available datasets, we conducted a controlled user study to demonstrate the effectiveness of our approach; the results indicate that users exposed to counterfactual visualizations formed more careful judgments about feature-to-outcome relationships.%
} %


\CCScatlist{ %
 \CCScat{K.6.1}{Management of Computing and Information Systems}%
{Project and People Management}{Life Cycle};
 \CCScat{K.7.m}{The Computing Profession}{Miscellaneous}{Ethics}
}

\teaser{
  \centering
  \includegraphics[width=1.0\linewidth]{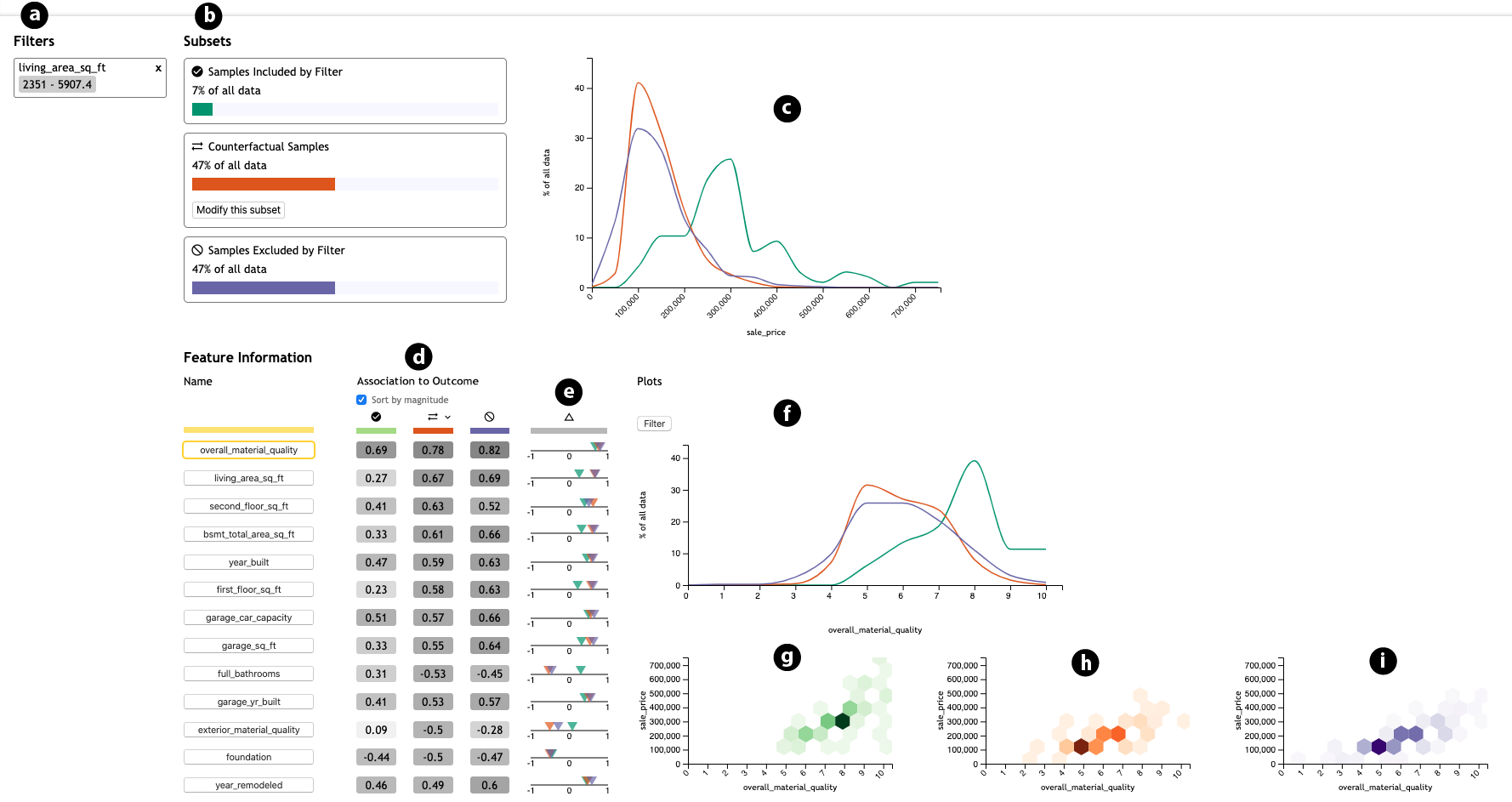}
  \caption{\label{analysis:main} Counterfactual visualization tool. The user has applied a filter constraint (a) to a multidimensional dataset and is shown the resulting included, counterfactual, and excluded subsets (b), their outcome distributions (c), and other feature information (d--i).}
}

\begin{document}

\firstsection{Introduction}

\maketitle

Supporting user inference and decision-making is one of the primary goals of information visualization, 
and data visualization systems therefore offer a variety of intuitive data representations and interactive tools to assist users. Visualization is used in a wide range of domains to explore problems and make decisions using large, complex, and high-dimensional datasets \cite{pak_chung_wong_visual_2004}, with some level of trust that it will reveal important, otherwise easy-to-miss information. 
Visualizations do provide crucial insights, but visual data analysis \hl{can often overlook the hidden influence of confounding variables.
This is especially true if users apply ad hoc filtering operations to visualize specific subsets of high-dimensional data \cite{borland_contextual_2018}. This ``overview first, zoom and filter, then details on demand'' workflow \cite{shneiderman_eyes_1996} is found in many popular data visualization tools, such as Tableau. Although this approach is invaluable for managing large data,  
it may} also 
encourage mistaken assumptions about causality or the strength of relationships between features. 
For example, imagine two groups of individuals, where group A is active on social media and group B is not. Presented only with visual information that shows that group A is overall unhappier, one might infer that social media activity determines individual levels of happiness.
While an analyst may conduct more rigorous data analyses to reach any definite conclusions, the absence of visual cues about the effect of other attributes on the two groups can lead to unconscious assumptions about social media's relationship with happiness, which can bias subsequent analytical work.
In this scenario, 
the theory of 
counterfactual thinking would urge the following questions: Could individuals be unhappy even without social media? What are other factors that contribute to unhappiness, more or less so than social media usage does? 

Fields such as causal analysis and explainable artificial intelligence have growing analytical and visual support for counterfactual thinking, but there is little to no such support for it in data exploration stages that precede causal analysis or data modeling via machine learning. We believe that confounding factors are easy to overlook during visual data exploration, as confirmed by the results presented in Section \ref{results}.
Motivated to fill this gap, we developed CoFact, a novel visual approach and system prototype that can reveal the presence of confounding factors during earlier stages of visual data analysis. CoFact enables users to interactively explore data, perform filtering operations, and analyze \textit{counterfactual subsets} to help guard against potentially erroneous conclusions about feature-to-outcome relationships.
The key contributions in this paper are summarized as follows:

\vspace{-0.15cm}
\begin{itemize}
    \item \textbf{Approach:} We present a novel way of visualizing and analyzing counterfactual subsets to investigate the influence of confounding factors within large and complex datasets.
 \vspace{-0.15cm}
    \item \textbf{Prototype:} We introduce CoFact, a visualization system prototype to reveal the influence of confounding variables and improve user decision-making during earlier stages of data analysis.
 \vspace{-0.15cm}    
    \item \textbf{Evaluation:} We demonstrate the effectiveness of our approach as implemented within CoFact through a controlled user study and interviews with 30 participants. Results show that the counterfactual visualizations significantly influenced user inference, reducing user confidence in weak feature-to-outcome relationships while confirming higher confidence in stronger feature-to-outcome relationships.
\end{itemize}
 \vspace{-0.15cm}    
\section{Related Work}
\vspace{-0.1cm}

The counterfactual-based approach to visualization presented in this paper builds upon prior research in a number of closely-related areas. This section discusses work in the areas of causal analysis, explainable machine learning, and data subset creation and analysis. 

\subsection{Counterfactual Thinking for Causal Analysis}

Early work on causality theory includes that by Pearl \cite{pearl_causality_2000, pearl_causal_1995}, Spirtes \cite{spirtes_causation_2000, spirtes_causality_1990}, and others. Causality theory rests on counterfactual thinking \cite{hume_treatise_1978, lewis_counterfactuals_2013}: if A causes B, then in an alternative, ``counterfactual'' scenario where A does not occur, B will not occur. Counterfactual thinking also asks us to investigate possible scenarios in which A does not occur but B occurs nonetheless. Byrne \cite{byrne_counterfactuals_2019} adds that counterfactuals can amplify causal judgement, since knowing that an alternative scenario that eliminates A would not lead to B would amplify one’s judgement of a causal relationship between A and B. Knowing that an alternative scenario eliminates A but also leads to B would weaken confidence in the influence of A on B. We revisit this idea in Section~\ref{def} to explain the value of visualizing counterfactual subsets. 

The use of large-scale data for causal inference and analysis is widespread, as researchers and analysts across domains seek to understand how data attributes influence certain outcomes \cite{wang_visual_2016}. Many approaches exist to identify and model causal relationships in observational data \cite{malinsky_causal_2018, pearl_causal_2009, shimizu_lingam_2014}, and several visualization systems support such analyses with suites of data visualization and interaction tools. Traditional visualizations include directed acyclic graph (DAG) layouts and Hasse diagrams \cite{jin_visual_2020}. Researchers have also proposed alternatives, such as Growing Squares \cite{elmqvist_growing_2003} and Growing Polygons \cite{elmqvist_causality_2003}, to enhance typical DAGs. The Visual Causality Analyst offers 2D graph views and statistical parameters to reveal possible causal influences of variables \cite{wang_visual_2016}. Others have introduced animations illustrating causal relationships \cite{kadaba_visualizing_2007} and a visual causal analysis system for hypothesis generation and evaluation \cite{chen_data_2011}. While determining causality is not this work's primary goal, casual analysis does provide the context within which CoFact aims to provide additional insight. In contrast to past approaches, CoFact does not depend on or depict abstract DAGs. Instead, it visualizes counterfactual subsets, as described later, to facilitate analysis during data exploration.

\subsection{Counterfactual Explanations in Machine Learning}
\vspace{-0.05cm}

Counterfactual thinking has gained increased recent attention in explainable machine learning research. Similar to counterfactual thinking for causal analysis \cite{lewis_counterfactuals_2013}, counterfactual explanations explore what modifications in the data would lead to an alternative prediction by a machine learning model \cite{wachter_counterfactual_2017, miller_explanation_2017}. Several techniques exist for generating counterfactual explanations \cite{ghazimatin_prince_2019, madumal_explainable_2019, liu_towards_2017, hohman_visual_2018, spinner_explainer_2019, karimi_model-agnostic_2019}. One example is DATE, which focuses on optimization and tree-based models \cite{lucic_actionable_2019}. Most techniques for counterfactual explanation generation focus on deep neural networks \cite{liu_towards_2016, strobelt_lstmvis_2016}, while LIME and DiCE are model-agnostic tools \cite{mothilal_explaining_2019, ribeiro_why_2016}. 

Visualization systems for counterfactual explanations have also been developed to present actionable insights. For example, DECE supports comparison of subsets’ counterfactual examples \cite{cheng_dece_2020}. Other systems include Prospector \cite{krause_interacting_2016} and RuleMatrix \cite{ming_rulematrix_2018}, both of which are model-agnostic. The What-If Tool \cite{wexler_what-if_2020} enables interactive exploration of machine learning models to help users find the nearest ``counterfactual'' data point. 
ViCE \cite{gomez_vice_2020} visualizes the minimum modification required to change a model’s prediction, and \cite{goyal_counterfactual_2019} introduces counterfactual visual explanations for image data. Lastly, \cite{poyiadzi_face_2020} addresses shortcomings of counterfactual explanations and proposes FACE to find ``feasible paths'' between a subject’s current and desired states. While such systems exist to help explore counterfactual possibilities in machine learning, counterfactual visualization support for data analysis in pre-modelling stages is rare.

\vspace{-0.05cm}
\subsection{Data Subsets and Counterfactual Possibilities} \label{related:subsets}
\vspace{-0.05cm}

This paper's focus is on the data exploration stages that precede causal analysis and modeling, during which data filtering and subset creation are common preliminary steps to examine the influence of certain features on an outcome~\cite{may_guiding_2011, keim_visual_2001}.
Existing systems employ various interaction techniques to help users create and analyze data subsets \cite{keim_information_2002, ferreira_de_oliveira_visual_2003, yi_toward_2007}, while others have proposed algorithms for automated feature selection \cite{seo_rank-by-feature_2005, tatu_automated_2011}. Visualization methods often employ correlation analysis and use scatter plots and heat maps to display feature information. Researchers have also proposed a range of geometrically transformed displays \cite{cleveland_visualizing_1993,chernoff_use_1973,keim_designing_2000, keim_pixel_2002, keim_visdb_1994, johnson_tree-maps_1991, leblanc_exploring_1990, shneiderman_tree_1992} and radial visualization methods \cite{draper_survey_2009, diehl_uncovering_2010} for feature evaluation. An example is \cite{elmqvist_datameadow_2008}, which adapts the RadViz technique \cite{hoffman_dna_1997, hoffman_dimensional_1999} to project attributes onto a 2D space. Others offer related approaches to dimension reduction for feature set evaluation  \cite{ingram_dimstiller_2010, choo_ivisclassifier_2010, cheng_data_2016}. 

Researchers have noted the uncertainty and bias problems inherent in making inferences based upon visual representations of user-defined subsets \cite{artur_novel_2019}. Additionally, visual data exploration is often undertaken by analysts with little a priori knowledge about the data features, lending them more vulnerable to whatever algorithmic or statistical biases are visualized \cite{elmqvist_datameadow_2008}. While the tools mentioned above help users examine data features and their relationships, they lack explicit visualizations of counterfactual possibilities that might illuminate confounding factors. Thus, in this paper, we augment standard data subset creation, analysis, and visualization techniques by visualizing counterfactual subsets as described in Section~\ref{cofact}. The goal of visualizing counterfactual data is to protect users from spurious assumptions about causal relationships until causality can be analytically established during later stages of data analysis.

\vspace{-0.1cm}
\section{CoFact} \label{cofact}
\vspace{-0.1cm}

This section describes the proposed counterfactual approach to visual data analysis (Section~\ref{def}) and the visual interface of CoFact (Section~\ref{ui}).
Two usage scenarios that demonstrate the approach and interface are detailed in Section~\ref{scenarios}.

\subsection{Counterfactual Approach} \label{def}
The primary goal of CoFact is to caution users against false inferences about a data variable's causal effect. To this end, CoFact visually alerts users to possible confounding factors during data exploration, which typically arise when they perform ad hoc filtering operations to analyze relationships between data variables. Key conceptual components of this visualization system include the \textit{counterfactual subset} and \textit{filter strength} as described below.

\subsubsection{Counterfactual Subset}

Consider the social media example introduced in Section~1. Typically, upon applying a filter constraint such as \textit{``social media status = active''} on a dataset of individuals, two subsets are created: the included subset of individuals who match the filter criterion (people active on social media) and the excluded subset of individuals who do not match it (people inactive on social media).
We propose a third, counterfactual subset comprising data samples that do not match the filter constraint but are similar to the included subset in other ways. Filtering thus creates three distinct subsets: (1) \textbf{the included subset (IN)}, (2) \textbf{the counterfactual subset (CF)}, which does not match the filter constraint but is similar to the included subset across other features, and (3) \textbf{the excluded subset (EX)} which does not match the filter constraint and also differs substantially from the included subset across other features. 

In the social media example, subset CF would include \textit{inactive} individuals who are \textit{similar} to active individuals across other features such as age, time spent online, and overall well-being, whereas subset EX would include \textit{inactive} individuals who are also \textit{different} from active individuals across other features.
We determine CF and EX by computing the similarity, as measured by the Euclidean distance \cite{danielsson_euclidean_1980}, between samples \textit{a\textsubscript{0}\dots a\textsubscript{n-1}} in IN and samples \textit{b\textsubscript{0}\dots b\textsubscript{m-1}} that do not match the filter constraint. For categorical features, the distance between any two values is 1 if they are equal and 0 if they are not.
For each \textit{b\textsubscript{j}}, we calculate the similarity to IN as the mean of the normalized distances to each \textit{a\textsubscript{i}}.
As a default, which was used in our user study, we form CF using the top 50\% most similar samples, and EX using
the bottom 50\%. 
 \hl{Section~\ref{future} discusses other potential methods (e.g.,~alternative distance and similarity metrics, different subset sizes) for determining the counterfactual subset.}
 
\subsubsection{Characterizing Filters} \label{definefilters}
Suppose that the data analysis goal is to understand to what extent various data features contribute to an outcome of interest. To do so, one can apply different filter constraints and observe their impact on the outcome distributions of the IN, CF, and EX subsets --- $o_{IN}$, $o_{CF}$, and $o_{EX}$ respectively. This impact, evaluated based on the extent to which the filter constraint ($f$) differentiates $o_{IN}$, $o_{CF}$, and $o_{EX}$, exists on a spectrum. At one extreme, after applying a  filter constraint, $o_{IN}$ might differ significantly from both $o_{CF}$ and $o_{EX}$, even though IN and CF are similar across features other than $f$, whereas CF and EX are overall more dissimilar but share $f$. Such an effect suggests a notable influence of $f$ on the outcome. At the other extreme, after applying a filter constraint, $o_{IN}$ and $o_{CF}$ might be very similar despite the difference in $f$, while $o_{EX}$ differs significantly from both.
Such an effect suggests a negligible influence of $f$ on the outcome because $o_{CF}$ and $o_{EX}$ are different despite sharing $f$. Between these two extremes, a filter constraint’s impact might lead to less definitive insights.

Given this spectrum, we characterize filter constraints on a scale from 0 to 1, where 0 corresponds to no impact by $f$ (i.e., $o_{IN}$~$=$~$o_{CF}$) and 1 corresponds to a large impact (i.e., a maximal difference between $o_{IN}$ and $o_{CF}$). 
To measure this difference and characterize filters accordingly, the Hellinger distance \cite{hellinger} is used for categorical outcomes and the Kolmogorov-Smirnov test \cite{kstest} for numerical outcomes. 
To evaluate the proposed counterfactual approach within CoFact, we defined three categories of \emph{Filter Strength} --- \textbf{weak} $\leq$~0.40 $<$ \textbf{moderate} $<$~0.60 $\leq$ \textbf{strong} --- to characterize a filter constraint’s impact using these measures (Table~\ref{strength}). These thresholds are based on a commonly used categorization of effect size on a spectrum from weak to strong \cite{evans}. We conducted an empirical analysis to check that these thresholds are suitable for both the Hellinger and Kolmogorov-Smirnov test measures.

\definecolor{gainsboro}{rgb}{0.86, 0.86, 0.86}
\begin{table}[ht]
\centering
\renewcommand{\arraystretch}{1.3}%
\begin{tabular}{|p{1.8cm}|p{1.6cm}|p{2cm}|p{1.6cm}|}
    \hline
    Filter Strength &\cellcolor{gainsboro}\textbf{Weak} & \cellcolor{gainsboro}\textbf{Moderate} & \cellcolor{gainsboro}\textbf{Strong} \\
    \hline
    IN/CF \newline Difference \textit{d} & \multirow{2}{1.6cm}{0~$\leq$~\textit{d}~$\leq$~0.40} & \multirow{2}{2cm}{0.40~$<$~\textit{d}~$<$~0.60} & \multirow{2}{1.7cm}{0.60~$\leq$~\textit{d}~$\leq$~1} \\
    \hline
\end{tabular}
\caption{\label{strength}Filter strength thresholds based on IN/CF difference.}
\vspace{0.35cm}
\end{table}

Consider again the social media and happiness example. Suppose that, upon applying the filter ``\textit{social media status = active}'' ($f$), one finds that active individuals (IN) and similar inactive individuals (CF) both have high levels of unhappiness
($o_{IN}~\sim~o_{CF}$),
whereas inactive individuals dissimilar to active individuals (EX) have 
low levels of happiness (\{$o_{IN}$, $o_{CF}\}~\nsim~o_{EX}$). 
This filter constraint would be considered \textbf{weak}, for there is little difference between IN and CF (\textit{d}~$\leq$~0.40). On the other hand, suppose that, after applying $f$, one finds that active individuals (IN) have high levels of unhappiness ($o_{IN}~\nsim~\{o_{CF}$, $o_{EX}$\}), whereas inactive individuals both similar (CF) and dissimilar (EX) to active individuals have low levels of happiness ($o_{CF}~\sim~o_{EX}$). In this case the filter constraint would be considered \textbf{strong} (\textit{d}~$\geq$~0.60).

Among the filters used in our evaluation of CoFact that are listed in Table~\ref{filters}, $F_{VRS}$ is weak because its corresponding IN/CF difference measure is 0.33. In contrast, $F_{S}$ is strong because its corresponding measure is 0.69. $F_{UNEMP1}$ is moderate, with a corresponding measure of 0.47. If a filter involves applying more than one feature constraint (e.g., ``\textit{social media status = active}'' \& ``\textit{age~$<$~20}''), the IN/CF difference is calculated in the same way using the final IN, CF, and EX distributions after the compounded filters are applied, as with user study tasks $T_{Moderate1}$, $T_{Weak2}$, and $T_{Strong3}$ in Table~\ref{tasks}.

\begin{figure}
    \centering
    \includegraphics[height=8cm]{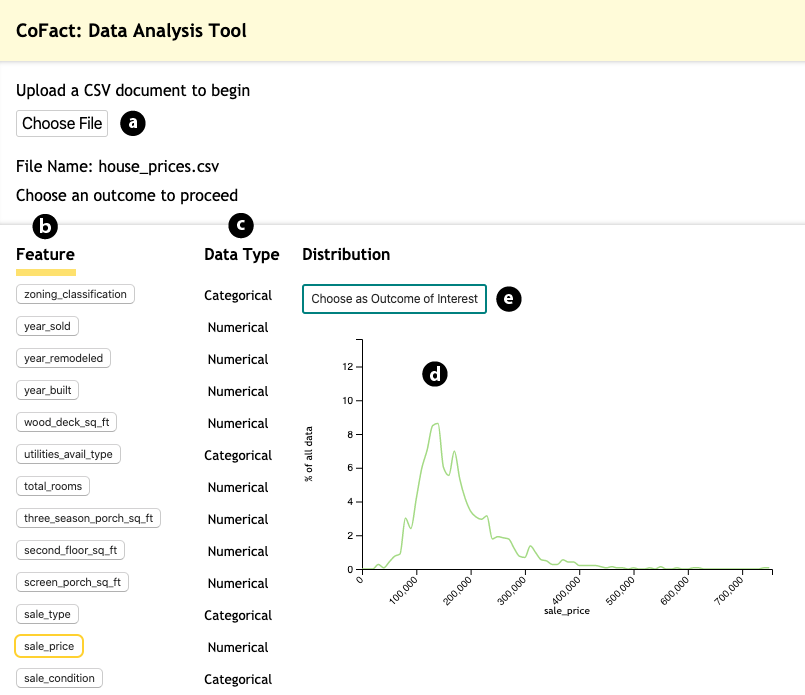} 
    \caption{\label{landing} CoFact landing page. The user must first choose a file (a) and then select an outcome of interest (b, e) to proceed.}
\end{figure}

\subsection{Design Requirements} \label{req}

Based on the motivation and purpose of CoFact, three key design requirements were identified. \hl{While not derived through a formal empirical study, these are informed by the authors’ long history of 
developing visual analytics tools in collaboration with experienced high-dimensional data users and analysts.}

 \vspace{-0.15cm}    
\begin{itemize}
    \item[\textbf{R1}] \textbf{Use feature information to choose, refine, and apply data filter constraints.} Feature statistics and visualizations should help users evaluate and choose filter constraints.
 \vspace{-0.15cm}    
    \item[\textbf{R2}] \textbf{Understand the included, counterfactual, and excluded subsets, and, relatedly, evaluate feature-to-outcome relationships.} Users should be able to understand the three subsets that result from filtering operations. Information presented about the subsets should inform user inference about the influence of a filter contraint feature on the outcome of interest. 
 \vspace{-0.15cm}    
    \item[\textbf{R3}] \textbf{Compare feature differences across subsets.} The presentation of the differences in features 
    across subsets should further user understanding of the features and of their relationships with the outcome of interest.
 \vspace{-0.15cm}    
\end{itemize}

\subsection{Visual Interface} \label{ui}

\begin{figure*}[t]
    \centering
    \includegraphics[height=9.4cm]{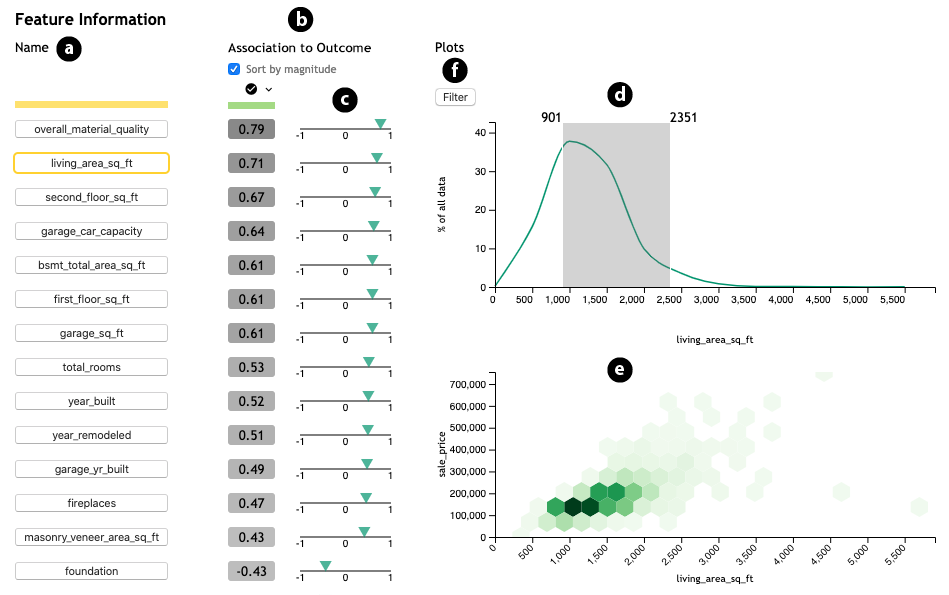}
    \vspace{-0.2cm}
    \caption{\label{feature:info} First analysis page. Feature information is displayed after the user chooses an outcome of interest to facilitate filter selection.}
    \vspace{-0.3cm}
\end{figure*}

We designed CoFact to improve user decision-making via the proposed counterfactual approach and to support design requirements \textbf{R1--R3}. CoFact is built as an Electron application using D3 and JavaScript, and it accepts as input CSV data tables in tidy format \cite{tidy-data}. 

\subsubsection{Initial Analysis Configuration}
The user begins by selecting a data file to load into CoFact. After loading, a summary of feature names and data types is displayed (Figure~\ref{landing}~(b--c)). A feature can be categorical (binary or non-dichotomous) or numerical. Clicking on any feature name button displays a corresponding distribution plot~(d). Distributions are shown using line charts for continuous variables and bar charts for categorical variables. The user then selects a feature to serve as the primary outcome of interest~(e). This outcome feature is used to compare subsets during the subsequent analysis.

After selecting an outcome of interest, the user is shown the first
analysis page (Figure~\ref{feature:info}). The primary analytical task is to explore how the outcome feature is influenced by other features. \hl{We thus provide information about feature-to-outcome associations. The association measures described next are intended to guide users, however they are not key components of the proposed counterfactual approach. Provided both the association and counterfactual information, users may engage in an iterative process in which they compare different sets of information and modify their analyses accordingly. 
The visualizations in CoFact present one specific implementation, however the general counterfactual approach is not limited to it.}

A list of feature buttons is displayed in the ``Name'' column~(a), which users can sort alphabetically in ascending or descending order. The ``Association to Outcome'' column~(b) shows how strongly a feature is associated with the outcome on a [-1, 1] or [0, 1] scale depending on the feature types.
If the feature and outcome are both numerical, this value is the Pearson correlation coefficient. If the feature or outcome is categorical and the other is numerical, multiple regression R\textsuperscript2 is used. If both the feature and outcome are categorical, Cramér's V is used to determine the association-to-outcome value. In our user study (Section~\ref{study}), participants viewed only correlation measures for simplicity.
Plots to the right of each association value graph it on a number line~(c). Users can click the green bar above the association-to-outcome column to sort in ascending or descending order. The ``Sort by magnitude'' checkbox enables users to ignore the signs of any correlation values.

The next analytical step is to choose a filter constraint to narrow the focus of the analysis. When a feature button~(a) is clicked, two plots display detailed information about the selected feature. The first plot~(d) shows the feature's distribution. 
Users can click and drag to select a range of feature values to use as a filter constraint. A line chart is used for numerical features and a bar chart 
for categorical features.
The second plot shows the selected feature's relationship to the outcome of interest (Figure~\ref{feature:info}~(e)). When the selected feature is categorical and the outcome is numerical, we use a violin plot.
When both the selected feature and the outcome are categorical, we use a heat map.
When both the selected feature and outcome are numerical, a hex map is used as shown in Figure~\ref{feature:info}~(e).

These visualizations help users decide upon a filter (\textbf{R1}). 
After selecting a feature and choosing the desired value via the distribution plot, users can click the ``Filter'' button (Figure~\ref{feature:info}~(f)) to perform a filtering operation.

\subsubsection{Subset Visualization}

\definecolor{burntorange}{rgb}{0.8, 0.33, 0.0}
\definecolor{green(munsell)}{rgb}{0.0, 0.66, 0.47}
\definecolor{royalpurple}{rgb}{0.47, 0.32, 0.66}

Applying a filter leads to the main analysis page displayed in Figure~\ref{analysis:main}. The filter constraint is listed in the ``Filters'' column~(a). The ``Subsets'' section~(b) on the right shows the three subsets created by filtering: the included (IN) \tikz\draw[green(munsell),fill=green(munsell)] (0,0) circle (.5ex);, counterfactual (CF) \tikz\draw[burntorange,fill=burntorange] (0,0) circle (.5ex);, and excluded (EX) \tikz\draw[royalpurple,fill=royalpurple] (0,0) circle (.5ex); subsets. The colored bars show what percentage of all data makes up the given subset. 
Plot~(c) shows the outcome distributions of each of the three subsets. 
For example, the orange line shows what percentage of CF has certain outcome values. These visualizations support \textbf{R2} by enabling users to compare outcome distributions across subsets to evaluate the applied filter's strength with respect to the outcome.

\subsubsection{Comparative Feature Information}

In Figure~\ref{analysis:main}~(d), ``Association to Outcome'' columns now display association values for IN, CF, and EX separately. 
The user can sort each column in ascending or descending order. 
Column~(e) graphs the association values for each subset on a number line. Plot~(f) graphs the selected feature's distributions for each of the three subsets. Below this graph, there are three separate feature vs. outcome plots for
\hl{IN~(g), CF~(h), and EX~(i).}
Taken together, the feature information and visualizations provided for each individual subset help users further explore features, edit existing filter constraints, and add new ones to the list of filters~(a). These components support \textbf{R3}. 

\subsection{Usage Scenarios} \label{scenarios}

\begin{figure}
    \centering
    \begin{subfigure}{0.23\textwidth}
        \includegraphics[width=\textwidth, height=2.7cm]{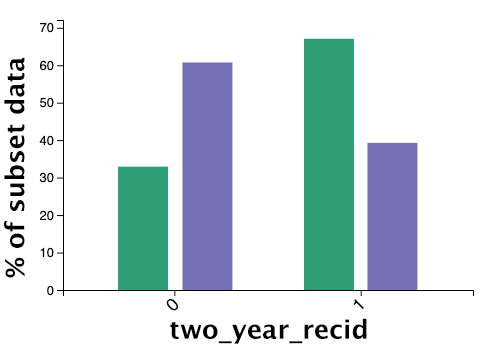}

        \caption{\centering Only IN \tikz\draw[green(munsell),fill=green(munsell)] (0,0) circle (.5ex); and EX$_{control}$ \tikz\draw[royalpurple,fill=royalpurple] (0,0) circle (.5ex);}
        \label{weak:sub1}
    \end{subfigure}
    \begin{subfigure}{0.23\textwidth}
        \includegraphics[width=\textwidth, height=2.7cm]{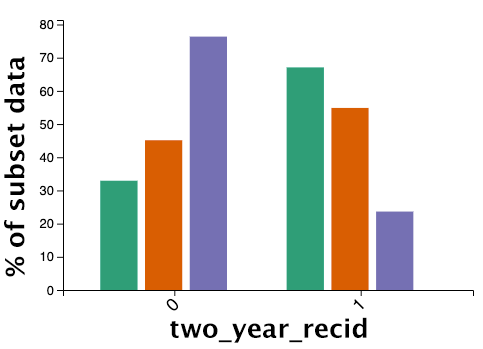}

        \caption{\centering IN \tikz\draw[green(munsell),fill=green(munsell)] (0,0) circle (.5ex);, CF \tikz\draw[burntorange,fill=burntorange] (0,0) circle (.5ex);, and EX \tikz\draw[royalpurple,fill=royalpurple] (0,0) circle (.5ex);}
        \label{weak:sub2}
    \end{subfigure}
    \caption{Subset distributions for the outcome (\textit{two\_year\_recid}) with a weak filter (\textit{6~$<$~v\_decile\_score~$<$~10}).}
    \label{weak}
    \begin{subfigure}{0.23\textwidth}
        \includegraphics[width=\textwidth, height=2.7cm]{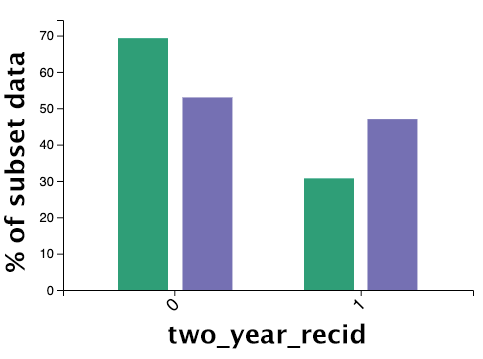} 
        \caption{\centering Only IN \tikz\draw[green(munsell),fill=green(munsell)] (0,0) circle (.5ex); and EX$_{control}$ \tikz\draw[royalpurple,fill=royalpurple] (0,0) circle (.5ex);}
        \label{strong:sub1}
    \end{subfigure}
    \begin{subfigure}{0.23\textwidth}
        \includegraphics[width=\textwidth, height=2.7cm]{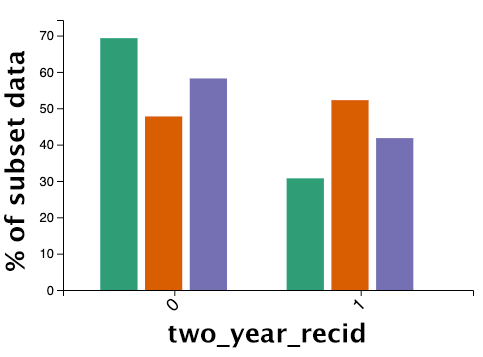}
        \caption{\centering IN \tikz\draw[green(munsell),fill=green(munsell)] (0,0) circle (.5ex);, CF \tikz\draw[burntorange,fill=burntorange] (0,0) circle (.5ex);, and EX \tikz\draw[royalpurple,fill=royalpurple] (0,0) circle (.5ex);}
        \label{strong:sub2}
    \end{subfigure}
    \caption{Subset distributions for the outcome (\textit{two\_year\_recid}) with a strong filter (\textit{sex~$=$~female}).}
    \label{strong}
\end{figure}

\begin{figure}
    \centering
    \begin{subfigure}{0.23\textwidth}
        \includegraphics[width=\textwidth, height=2.7cm]{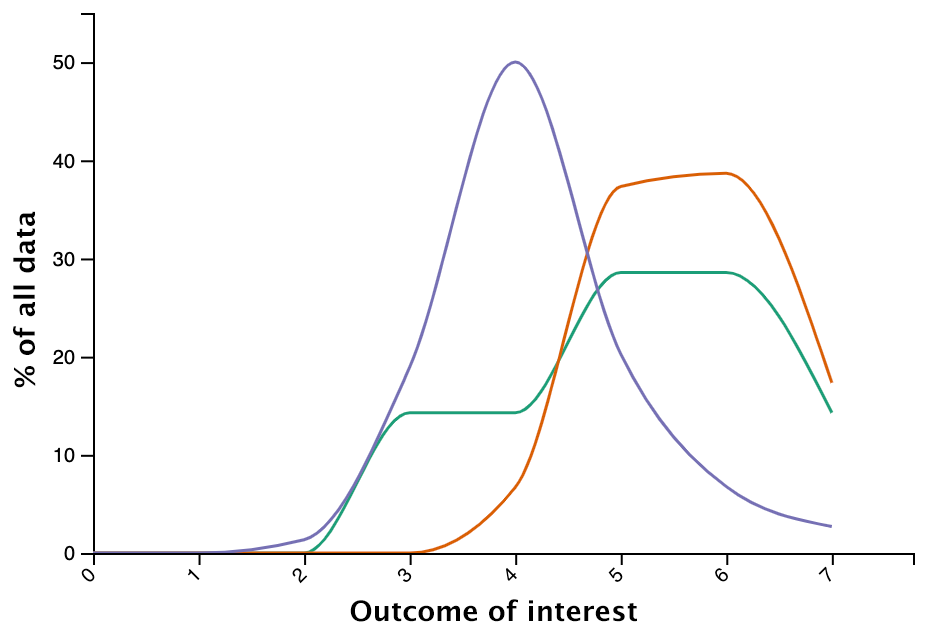} 
        \caption{\centering IN \tikz\draw[green(munsell),fill=green(munsell)] (0,0) circle (.5ex);, CF \tikz\draw[burntorange,fill=burntorange] (0,0) circle (.5ex);, EX \tikz\draw[royalpurple,fill=royalpurple] (0,0) circle (.5ex); with weak filter}
        \label{weakquant}
    \end{subfigure}
    \begin{subfigure}{0.23\textwidth}
        \includegraphics[width=\textwidth, height=2.7cm]{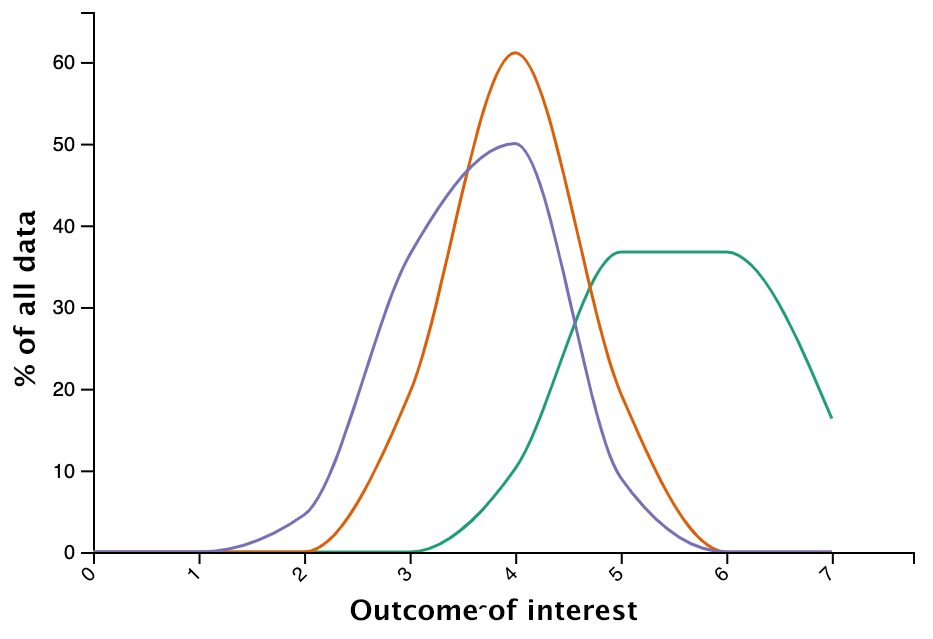}
        \caption{\centering IN \tikz\draw[green(munsell),fill=green(munsell)] (0,0) circle (.5ex);, CF \tikz\draw[burntorange,fill=burntorange] (0,0) circle (.5ex);, and EX \tikz\draw[royalpurple,fill=royalpurple] (0,0) circle (.5ex); with strong filter}
        \label{strongquant}
    \end{subfigure}
    \caption{Example subset distributions for a numerical outcome.}
    \label{quant}
\end{figure}

This section describes
two usage scenarios
to illustrate the concepts from Section~\ref{def} using the visual interface introduced in Section~\ref{ui}. 
User study participants engaged with these and six other scenarios as described in Section~\ref{study}. These two scenarios use a dataset which includes information related to criminal recidivism \cite{recidivism}. The outcome of interest is \emph{Recidivism Within Two Years} (\emph{two\_year\_recid}), a categorical feature which the user selects on the landing page (Figure~\ref{landing}). A value of 0 for \emph{two\_year\_recid} indicates that the defendant was not arrested again for an offense within two years since the last arrest. A value of 1 indicates that they were arrested again within two years.
Using the information and visualizations on the first analysis page (Figure~\ref{feature:info}), the user can now choose and apply a filter constraint.

\definecolor{burntorange}{rgb}{0.8, 0.33, 0.0}
\definecolor{green(munsell)}{rgb}{0.0, 0.66, 0.47}
\definecolor{royalpurple}{rgb}{0.47, 0.32, 0.66}

\subsubsection{Weak Filter Scenario} \label{scenario_weak}
The user first analyzes the influence of \emph{Violent Recidivism Risk} (\textit{v\_decile\_score}) on \emph{two\_year\_recid}, the outcome of interest. To narrow their focus, they apply a filter constraint that \textit{v\_decile\_score} should be \textit{between 6 and 10}, which are relatively high risk scores. This operation creates distinct subsets to be visualized, similar to Figure~\ref{analysis:main} (which uses a numerical outcome). Figure~\ref{weak} displays the outcome distributions for each subset. Figure~\ref{weak}~(a) only visualizes defendants with high risk scores~\tikz\draw[green(munsell),fill=green(munsell)] (0,0) circle (.5ex); and those with low risk scores~\tikz\draw[royalpurple,fill=royalpurple] (0,0) circle (.5ex);; the counterfactual subset is empty. \hl{These visualized subsets are denoted IN and EX$_{control}$, respectively. Figure~\ref{weak}~(b), on the other hand, includes a non-empty counterfactual subset (CF~\tikz\draw[burntorange,fill=burntorange] (0,0) circle (.5ex);), and EX~\tikz\draw[royalpurple,fill=royalpurple] (0,0) circle (.5ex); is distinct from CF. IN~\tikz\draw[green(munsell),fill=green(munsell)] (0,0) circle (.5ex); is identical in (a) and (b).} Looking only at plot (a), the user is likely to have high confidence in \textit{v\_decile\_score}'s influence on \emph{two\_year\_recid}, as there is a significant visual difference in the outcome distributions of the two subsets.

Figure~\ref{weak}~(b)
offers contradicting information. Here, CF comprises defendants with low scores who are similar to IN in other ways. Individuals in EX still have low recidivism scores but are also more different from IN across other features than are individuals in CF. This plot shows that CF, despite having low scores, has a similar outcome distribution as IN (defendants with high scores) and a visually distinct outcome distribution from EX. This visualization alerts the user to possible confounding features that influence the outcome more strongly than \textit{v\_decile\_score}, leading them to lower their confidence in the influence of \textit{v\_decile\_score} on \emph{two\_year\_recid}. 

\hl{For visual reference, Figure~\ref{quant}~(a) shows what a weak filter scenario may look like for a different numerical outcome of interest. As with Figure~\ref{weak}~(b), IN~\tikz\draw[green(munsell),fill=green(munsell)] (0,0) circle (.5ex); and CF~\tikz\draw[burntorange,fill=burntorange] (0,0) circle (.5ex); have similar outcome distributions, and both are different from the outcome distribution of EX~\tikz\draw[royalpurple,fill=royalpurple] (0,0) circle (.5ex);.}

\subsubsection{Strong Filter Scenario} \label{scenario_strong} 
Next, the user analyzes the influence of \emph{sex} on \emph{two\_year\_recid}. To narrow their focus, they apply a filter constraint that \textit{sex} should be \textit{female}. This operation again creates distinct subsets for visualization. Figure~\ref{strong} displays the outcome distributions for each subset. Figure~\ref{strong}~(a) only visualizes female \hl{(IN)}~\tikz\draw[green(munsell),fill=green(munsell)] (0,0) circle (.5ex); and male \hl{(EX$_{control}$)}~\tikz\draw[royalpurple,fill=royalpurple] (0,0) circle (.5ex); defendants. Looking only at this plot, the user is likely to again have relatively high confidence that \textit{sex} influences \emph{two\_year\_recid}, albeit somewhat lower than they did for \textit{v\_decile\_score} due to a less stark difference in the subsets' outcome distributions.

In this scenario, Figure~\ref{strong}~(b) shows CF~\tikz\draw[burntorange,fill=burntorange] (0,0) circle (.5ex);, comprising male defendants who are similar to female defendants (IN \tikz\draw[green(munsell),fill=green(munsell)] (0,0) circle (.5ex);) in other ways. EX~\tikz\draw[royalpurple,fill=royalpurple] (0,0) circle (.5ex); are male defendants who are more dissimilar to IN. This plot shows that CF, despite being similar to IN across other features, has a noticeably different outcome distribution than IN. CF's outcome distribution is actually close to that of EX. Thus, despite other similarities, the difference in sex differentiates outcome distributions for the female (IN) and male (CF,~EX) subsets. This confirms or increases the user's high confidence in the influence of \textit{sex} on \emph{two\_year\_recid}.

\hl{Figure~\ref{quant}~(b) shows an example strong filter scenario for a numerical outcome. Similar to Figure \ref{strong} (b), IN~\tikz\draw[green(munsell),fill=green(munsell)] (0,0) circle (.5ex); has a distinctly different outcome distribution than both CF~~\tikz\draw[burntorange,fill=burntorange] (0,0) circle (.5ex); and EX~~\tikz\draw[royalpurple,fill=royalpurple] (0,0) circle (.5ex);, which are similar to each other.}
\section{User Study} \label{study}

\definecolor{gainsboro}{rgb}{0.86, 0.86, 0.86}
\begin{table*}[ht]
\centering
\renewcommand{\arraystretch}{1.3}%
\begin{tabular}{|p{1.2cm}|p{2.6cm}|p{6.8cm}|p{3.4cm}|}
    \hline
    \textbf{Filter} & \textbf{Strength \newline (IN/CF Difference)} & \textbf{Description} & \textbf{Constraint} \\
    \hline
    \multicolumn{4}{|c|}{\cellcolor{gainsboro}\textbf{Dataset 1:} Housing Prices, \textbf{Outcome of Interest:} Sale Price} \\
    \hline
    $F_{LA}$ & Strong (0.60) & Living area square footage & Within range 2200--5100 \\
    \hline
    \multicolumn{4}{|c|}{\cellcolor{gainsboro}\textbf{Dataset 2:} COVID-19 State Policies, \textbf{Outcome of Interest:} Percentage of State Population with Positive Cases} \\
    \hline
    $F_{SIP}$ & Weak (0.18) & Date state started shelter in place & On a day in April \\
    $F_{UNEMP1}$ & Moderate (0.47) & Percentage of state population unemployed & Within range 1--3.5 \\
    $F_{RNEB}$ & Weak (0.25) & Date state reopened non-essential businesses & On a day in March or April \\
    \hline
    $F_{WUIM}$ & Strong (0.68) & Weekly unemployment insurance maximum provided & Within range 240--350 \\
    $F_{UNEMP2}$ & Weak (0.38) & Percentage of state population unemployed & Equal to or above 5 \\
    \hline
    \multicolumn{4}{|c|}{\cellcolor{gainsboro}\textbf{Dataset 3:} Recidivism, \textbf{Outcome of Interest:} Recidivism within Two Years} \\
    \hline
    $F_{VRS}$ & Weak (0.33) & Violent recidivism risk & Within range 6--10 \\
    \hline
    $F_{S}$ & Strong (0.69) & Sex & Female \\
    \hline
\end{tabular}
\vspace{-0.15cm}
\caption{\label{filters}Filter constraints for the experimental data analysis tasks listed in Table \ref{tasks}.}
\end{table*}

\definecolor{gainsboro}{rgb}{0.86, 0.86, 0.86}
\begin{table*}[ht]
\centering
\renewcommand{\arraystretch}{1.3}%
\begin{tabular}{|p{2.6cm}|p{1.3cm}|p{.9cm}|p{1.4cm}|p{1cm}|p{1.1cm}|p{1.3cm}|p{1.1cm}|p{1.1cm}|}
    \hline
    &\cellcolor{gainsboro}\textbf{Dataset 1} &
    \multicolumn{5}{|c|}{\cellcolor{gainsboro}\textbf{Dataset 2}}
    &
    \multicolumn{2}{|c|}{\cellcolor{gainsboro}\textbf{Dataset 3}} \\
    \hline
    \textbf{Task} & $T_{Strong1}$ & $T_{Weak1}$ & $T_{Moderate1}$ & $T_{Weak2}$ & $T_{Strong2}$ & $T_{Strong3}$ & $T_{Weak3}$ & $T_{Strong4}$ \\
    \hline
    \textbf{Task Strength \newline (IN/CF Difference)} & Strong (0.60) & Weak (0.18) & Moderate (0.43) & Weak (0.20) & Strong (0.68) & Strong (0.82) & Weak (0.33) & Strong (0.69) \\
    \hline
    \textbf{Filter(s) Applied} & $F_{LA}$ &
    $F_{SIP}$ & 
    $F_{SIP}$ + \newline $F_{UNEMP1}$ & 
    $F_{SIP}$ + \newline $F_{RNEB}$ & 
    $F_{WUIM}$ & 
    $F_{WUIM}$ + \newline $F_{UNEMP2}$ & 
    $F_{VRS}$ & $F_{S}$ \\
    \hline
\end{tabular}
\vspace{-0.15cm}
\caption{\label{tasks}User study experimental data analysis tasks, using the filter constraints listed in Table \ref{filters}.}
\end{table*}

This section describes a controlled user study (n~$=$~30) that evaluates CoFact's ability to convey counterfactual possibilities. Results from the experimental tasks and feedback from post-study interviews (Section~\ref{results}) suggest that the counterfactual visualizations improved user inference about feature-to-outcome relationships without hampering system usability.

\subsection{Hypotheses} \label{hypotheses}
The user study was designed to test the following hypotheses:

\textbf{Hypothesis 1.} Users exposed to the counterfactual visualizations will have lower confidence in the influence of a weak filter on the outcome of interest than users who do not view the counterfactual subset. Relatedly, users exposed to the counterfactual visualizations will have a similarly high confidence in the influence of a strong filter on the outcome of interest as users who do not view the counterfactual subset.

\textbf{Hypothesis 2.} Upon being exposed to the counterfactual subset, users who initially did not view the counterfactual visualizations will have decreased confidence in a weak filter's influence but will maintain relatively high confidence in a strong filter's influence.

\textbf{Hypothesis 3.} The added counterfactual subset visualization capability will not significantly decrease the system's usability, efficiency, or the overall quality of user experience.

Users should have decreased confidence in a weak filter after viewing the counterfactual subset because this subset reveals the presence of counterfactual explanations for differences in outcome distributions that lie beyond the difference due to the filter constraint. Users should maintain relatively high confidence in a strong filter because the resulting counterfactual subset’s distribution would be distinct from that of the included subset, suggesting that the filter variable is valuable in explaining differences in outcome.

\subsection{Data} \label{data}
We used three publicly available datasets for the user study. Dataset~1 was obtained from Kaggle and includes information (163 features, n~$=$~1500) about houses and their sale prices \cite{housing}. Dataset~2 contains information (42 features, n~$=$~50) related to the COVID-19 pandemic. It was formed using two separate publicly available datasets: (1)~information about COVID-19 cases and deaths in U.S. states \cite{covid1} and (2)~U.S. state policies related to the pandemic \cite{covid2}. Dataset~3, published by \textit{ProPublica}, contains information (20 features, n~$=$~1500) related to criminal recidivism \cite{recidivism}. 

\subsection{Design and Procedure}
\subsubsection{Participants and Groups}
We recruited 30 participants (male~$=$~14, female~$=$~16) via a campus-wide email, department mailing lists, and recruitment efforts within our professional network. All participants were at least 18 years old and were either pursuing or had attained a university degree. Participants belonged to a diverse range of academic and professional sectors. Under a between-subjects design, 15 were randomly assigned to the control group (C) and primarily saw visualizations for only two subsets, while 15 were assigned to the counterfactual group (CF) and viewed the counterfactual subset as well. 

\definecolor{orange(colorwheel)}{rgb}{1.0, 0.5, 0.0}
\definecolor{azure(colorwheel)}{rgb}{0.0, 0.5, 1.0}
 
\subsubsection{Procedures and Study Tasks}

\begin{table*}[ht]
\centering
\renewcommand{\arraystretch}{1.3}%
\begin{tabular}{|p{2.4cm}|p{5.8cm}|p{5.8cm}|}
    \hline
    &\textbf{Counterfactual Group (CF)} & \textbf{Control Group (C)}\\
    \hline
    \textbf{[Q1] Difference \newline in outcome \newline distributions} &
    \textbf{Q1-CF1:} On a scale of 1 to 7, how different are the outcome distributions for the counterfactual and included subsets? \newline
    \newline \textbf{Q1-CF2:} On a scale of 1 to 7, how different are the outcome distributions for the counterfactual and excluded subsets? &
    \textbf{Q1-C:} On a scale of 1 to 7, how different are the outcome distributions for the included and excluded subsets? \\ 
    \hline
    \textbf{[Q2] Comments} &
    \multicolumn{2}{|l|}{What does this make you think about the filter variable's influence on the outcome?} \\
    \hline
    \textbf{[Q3] Confidence} &
    \multicolumn{2}{|l|}{On a scale of 1 to 7, how confident are you in the filter variable's influence on the outcome?} \\
    \hline
\end{tabular}
\caption{\label{questions} Experimental questions participants responded to after each data analysis task.}
\end{table*}

\begin{table*}[ht]
\centering
\renewcommand{\arraystretch}{1.3}%
\begin{tabular}{|p{3.2cm}|p{0.7cm}|p{10.1cm}|}
    \hline
    \textbf{Metric}&&\textbf{Criterion}\\
    \hline
    \textbf{[M1] Usability} &
    \textbf{M1.1} \newline
    \textbf{M1.2} \newline
    \textbf{M1.3} &
    Users should find the system overall easy to use. \newline
    It should be easy to add and remove filters. \newline
    It should be easy to understand the  visual subset divisions that result from a filter operation. \\
    \hline
    \textbf{[M2] Informativeness} &
    \textbf{M2.1} \newline\newline 
    \textbf{M2.2} \newline\newline
    \textbf{M2.3} \newline\newline 
    \textbf{M2.4} &
    Users should find the system informative overall and use it to explore data in a way they would not be able to otherwise. \newline
    Users should find the feature-to-outcome correlations and feature distributions informative. \newline
    Users should find the differences in feature-to-outcome correlations and feature distributions across subsets informative. \newline
    Users should find the counterfactual visualization capability informative and helpful for better exploring feature-to-outcome relationships. \\
    \hline
    \multicolumn{2}{|l|}{\textbf{[M3] Efficiency}} &
    Users should analyze and understand data more efficiently using CoFact. \\
    \hline
\end{tabular}
\caption{\label{questions:post} Criteria to evaluate user experience.}
\end{table*}

User study sessions were conducted remotely using Zoom video conferencing, and each lasted for roughly one hour. First, participants answered a pre-study questionnaire that asked them, among other questions, to rate their level of expertise on a scale from 1 (novice) to 7 (expert). Groups C and CF reported no significant difference in expertise for both general data analysis ($p=0.50$) and visual data analysis ($p=~0.37$).

Next, we reviewed essential terms \hl{(e.g., counterfactual) and 
gave participants a tour of the visual interface.}
Participants were then provided remote control of the moderator's screen and guided through some practice data analysis tasks in CoFact, \hl{before they completed the main experimental tasks} listed in Table~\ref{tasks}. Each task's strength and corresponding IN/CF Difference measure are provided in the second column according to the methodology described in Section~\ref{definefilters}. Each participant completed 7 analysis tasks, while 23 participants (14 in group C, 9 in group CF) also had time to complete an additional task (\bm{$T_{\bm{Weak2}}$}). Each task involved applying the corresponding filter constraints detailed in Table~\ref{filters} and then responding to questions \textbf{Q1--Q3} listed in Table~\ref{questions}.
The questions asked users to describe what they observed and what inferences they drew using the visualizations. 

After the pre-study questionnaire, group CF participants completed the experimental tasks. Group C participants did the same, but they also repeated \bm{$T_{\bm{Weak3}}$} and \bm{$T_{\bm{Strong4}}$} after being exposed to the counterfactual subset. Despite the time constraint, this variation for group C enabled us to gather within-subjects results for at least two tasks. Section~\ref{scenarios} describes the expected behaviors for \bm{$T_{\bm{Weak3}}$} and \bm{$T_{\bm{Strong4}}$} in more detail. Finally, participants provided post-study feedback about the tool's usefulness and their overall experience via a questionnaire.

\subsubsection{Measures}

After performing each of the tasks listed in Table~\ref{tasks}, participants reported their confidence in each filter constraint's influence on the outcome of interest on a scale from 1 (no confidence) to 7 (high confidence) in response to \textbf{Q3} in Table~\ref{questions}. For the post-study questionnaire, participants provided their level of agreement with eight statements related to the criteria listed in Table~\ref{questions:post}, on a scale from 1 (strongly disagree) to 7 (strongly agree).
\section{Results} \label{results}

This section reports results and feedback from the user study (Section~\ref{study}). Overall, results support \textbf{Hypotheses 1--3} (Section~\ref{hypotheses}), requirements \textbf{R1--R3} (Section~\ref{req}), and user experience metrics \textbf{M1--M3} (Table~\ref{questions:post}).

\subsection{Confidence}
We evaluated self-reported confidence with a 2 (treatment group: C~\tikz\draw[azure(colorwheel),fill=azure(colorwheel)] (0,0) circle (.5ex); vs. CF~\tikz\draw[orange(colorwheel),fill=orange(colorwheel)] (0,0) circle (.5ex);) $\times$ 2 (strength: weak vs. strong) repeated measures ANOVA with treatment group as a between-subjects factor and strength as a within-subjects factor using the afex package in R. See Table~\ref{tab:aov}. Post-hoc analysis was performed using estimated-marginal means with Tukey method adjustments for repeated tests. The \bm{$T_{\bm{Moderate1}}$} task was removed from this analysis due to there being only a single moderate task in the experiment.

\hl{Below we analyze differences in self-reported confidence between the C and CF treatment groups with respect to overall task strength and individual tasks. We report significant results with effect sizes and confidence intervals. In general the effect sizes are moderate to strong, which provides support for \textbf{Hypotheses 1 and 2} related to participant confidence.}

\begin{table}[ht]
\centering
\begin{tabular}{lrrr}
  \hline
 &   F & $\eta^2$ & $p$ \\ 
  \hline
TREATMENT  & 3.44 & 0.07 & 0.07 \\ 
  STRENGTH  & 39.33 & 0.35 & 0.00 \\ 
  TREATMENT:STRENGTH & 6.42 & 0.08 & 0.02 \\ 
   \hline
 \end{tabular}
\caption{Results from the 2 (treatment group: C vs. CF) $\times$ 2 (strength: weak,  strong) repeated measures ANOVA evaluating user confidence.}
\label{tab:aov}
\end{table}

A significant main effect of strength was found ($F(1, 28)=39.33$, $p<0.001$, $\eta^2 = 0.35$).
Participants were significantly more confident with the strong tasks ($M=5.15$, $SD=1.43$) compared to the weak tasks ($M=3.77$, $SD=1.67$). This main effect is quantified by the higher order significant treatment $\times$ strength interaction ($F(1, 28)=6.42$, $p=0.02$, $\eta^2 = 0.08$).

\begin{figure}
    \centering
    \includegraphics[width=.35\textwidth]{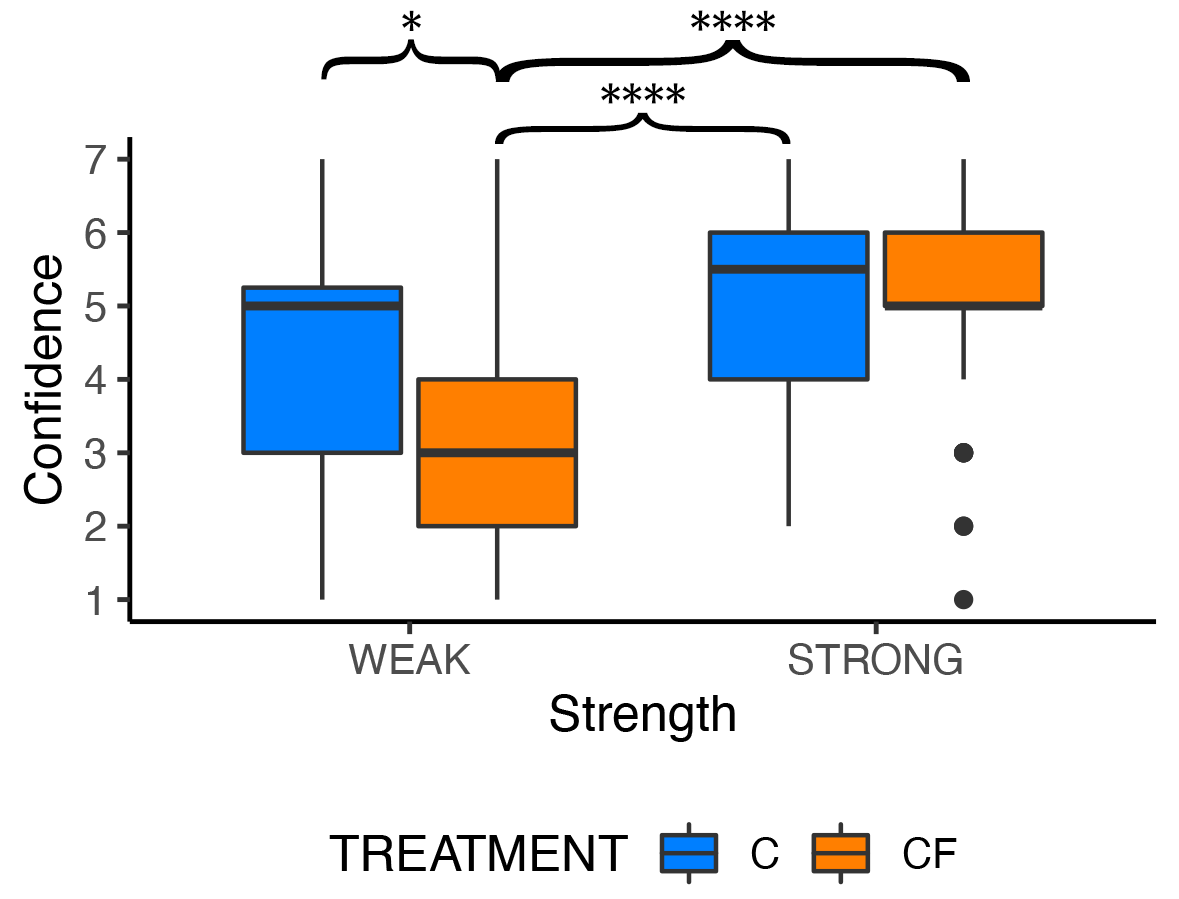} 
    \vspace{-0.3cm}
    \caption{Box plots of confidence for strong and weak tasks by treatment group.}
    \label{fig:confidence}
    \vspace{-0.35cm}
\end{figure}

 The CF treatment group was significantly less confident when evaluating weak tasks ($M=3.26$, $CI=[2.75,3.76]$) compared to all other conditions. The CF treatment group evaluating weak tasks was less confident than the CF treatment group  evaluating strong tasks ($M=5.17$, $CI=[4.67, 5.67]$, $t(28)=6.23$, $p<0.0001$), the C treatment group evaluating weak tasks ($M=4.32$, $CI=[3.82, 4.82]$, $t(52.8)=3.02$, $p=0.02$), and the C treatment group evaluating strong tasks ($M=4.63$, $CI=[4.63, 5.63]$, $t(52.8)=5.32$, $p<0.0001$). No other significant results were found. See Figure~\ref{fig:confidence}. 
 These results
support \textbf{Hypothesis~1}, that participants exposed to the counterfactual visualizations (CF) will have lower confidence in the influence of a weak filter on the outcome of interest than participants who do not view the counterfactual subset (C), but will have a similarly high confidence in the influence of a strong filter.

\begin{table}[ht]
\centering
\begin{tabular}{lrrr}
  \hline
&   F & $\eta^2$ & $p$ \\ 
  \hline
TREATMENT  & 1.33 & 0.02 & 0.26 \\ 
  TASK & 8.75 & 0.24 & 0.00 \\ 
  TREATMENT:TASK & 3.40 & 0.11 & 0.01 \\ 
   \hline
\end{tabular}
\caption{Results from the 2 (treatment group: C vs. CF) $\times$ 8 (task) repeated measures ANOVA evaluating user confidence for each task.}
    \label{tab:confidence_task}
\end{table}

\begin{figure*}
    \centering
    \includegraphics[width=.81\textwidth]{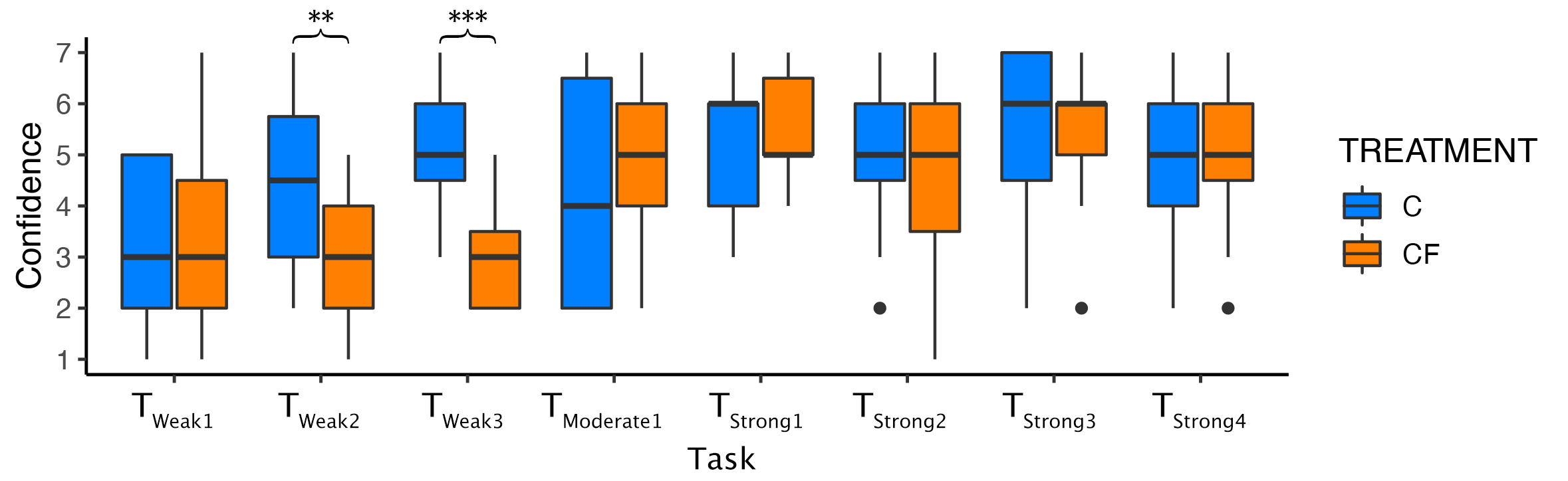} 
    \vspace{-0.10cm}
    \caption{Box plots of confidence by treatment group for each task.}
    \vspace{-0.06cm}
    \label{fig:confidence_trial}
\end{figure*}

We further investigated if there were differences between treatment groups for each of the tasks with a 2 (treatment group: C vs. CF) $\times$ 8 (task) repeated measures ANOVA (Table~\ref{tab:confidence_task}). Post-hoc analysis was again performed using estimated-marginal means with Tukey method adjustments for repeated tests.
A significant main effect of task was found ($F(4.40, 92.43)=8.75$, $p<0.001$, $\eta^2=0.24$). This main interaction was quantified by the higher-order significant treatment $\times$ task interaction ($F(4.40, 92.43)=3.40$, $p=0.01$, $\eta^2=0.11$). Post-hoc analysis was performed pair-wise comparing treatments for each task using estimated-marginal means with Tukey method adjustments for repeated tests. \hl{Significant differences were found between the C and CF treatments for \bm{$T_{\bm{Weak2}}$} ($t(144)=2.59$, $p=0.01$, $\eta^2 = 0.04$) and \bm{$T_{\bm{Weak3}}$} ($t(144)=3.76$, $p=0.0002$, $\eta^2 = 0.09$).} Participants in group CF had significantly lower confidence compared to those in group C. See Figure~\ref{fig:confidence_trial}. 

These results further support \textbf{Hypothesis~1}. The lone exception is for task \bm{$T_{\bm{Weak1}}$}, where participants in group CF did not have significantly lower confidence than participants in group C. A possible explanation for this phenomenon relates to existing knowledge of the subject matter. As information about the COVID-19 pandemic was widely spread at the time of the user study, participants may have had difficulty separating prior knowledge from the graphical presentation of the subset outcome distributions. The plotted lines may not have fit with their understanding of the real-world role of the filter variable ($F_{SIP}$ in Table~\ref{filters}) and thus confused their thinking. Participants often indicated that they did not know what to make of this scenario, and confidence was relatively low for both the C and CF groups (median value of 3).

As noted earlier, group C participants repeated a single weak task (\bm{$T_{\bm{Weak3}}$}) and strong task (\bm{$T_{\bm{Strong4}}$}) a second time with the counterfactual visualizations. We investigated the effect on confidence for these tasks and participants. Analysis was performed with a 2 (strength: weak vs. strong) $\times$ 2 (treatment group: C vs. CF) repeated-measures ANOVA with both strength and treatment as within-participant variables. 

\begin{figure}
    \centering
    \includegraphics[width=.34\textwidth]{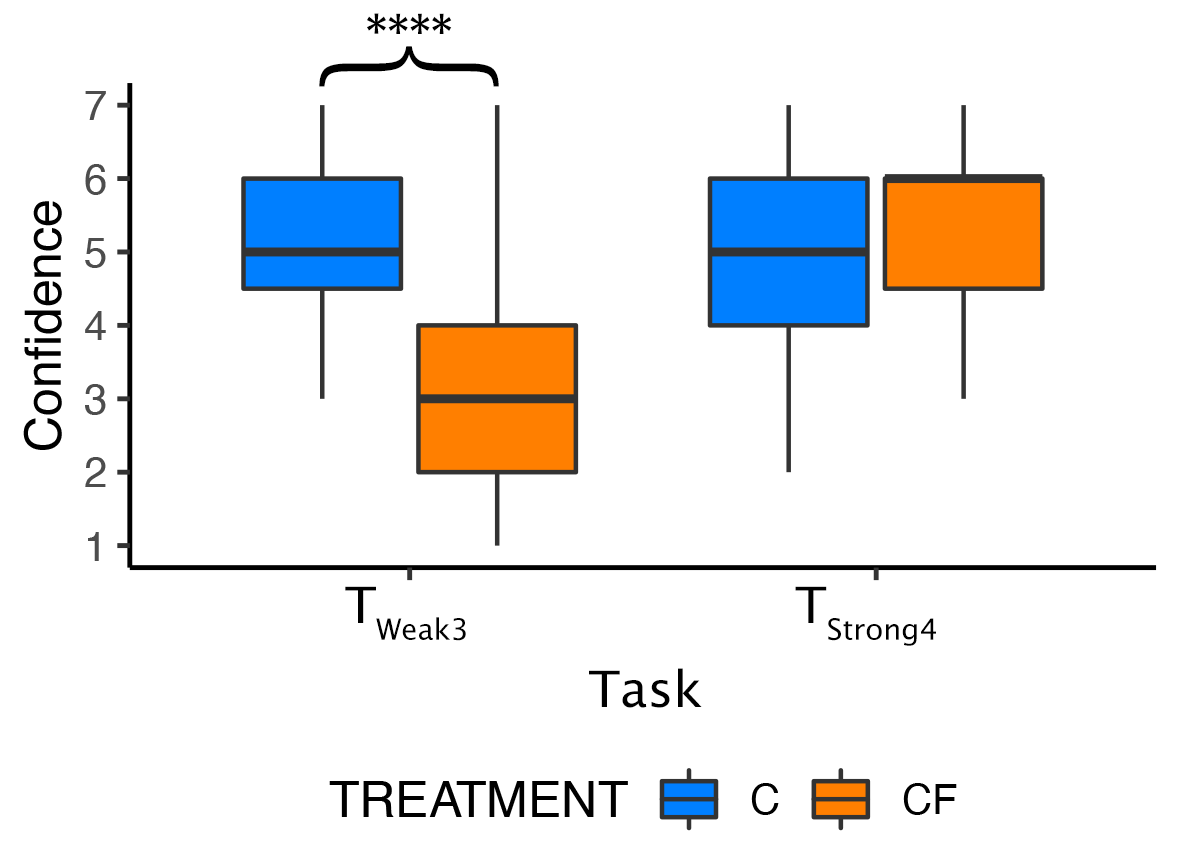} 
    \vspace{-0.08cm}
    \caption{Box plots of confidence for the \bm{$T_{\bm{Weak3}}$} and \bm{$T_{\bm{Strong4}}$} tasks for participants who started in the C condition and then repeated these tasks in the CF condition.}
    \label{fig:within}
\end{figure}

Significant main effects of strength ($F(1, 14)=8.73$, $p=0.01$, $\eta^2=0.08$) and treatment ($F(1, 14)=8.20$, $p=0.01$, $\eta^2=0.08$) were found. These main effects were quantified by the higher-order strength $\times$ treatment interaction  ($F(1, 14)=24.80$, $p<0.001$, $\eta^2=0.21$). Post-hoc analysis was performed comparing treatment for each strength. \hl{For \bm{$T_{\bm{Weak3}}$}, participants in the CF condition had significantly lower confidence ($M=3.13$, $SE=0.36$, $CI=[2.4, 3.87]$) compared to those in the C condition ($M=5.33$, $SE=0.36$, $CI=[4.6, 6.1]$, $t(28)=5.55$, $p<0.0001$). No significant difference was found between the C and CF conditions in \bm{$T_{\bm{Strong4}}$} ($t(28)=-1.51$, $p=0.14$).} See Figure~\ref{fig:within}. These results support \textbf{Hypothesis~2}, that user confidence would decrease for a weak filter but remain relatively similar and high for a strong filter upon viewing the counterfactual visualization.

\begin{figure}
    \centering
    \includegraphics[height=5.6cm]{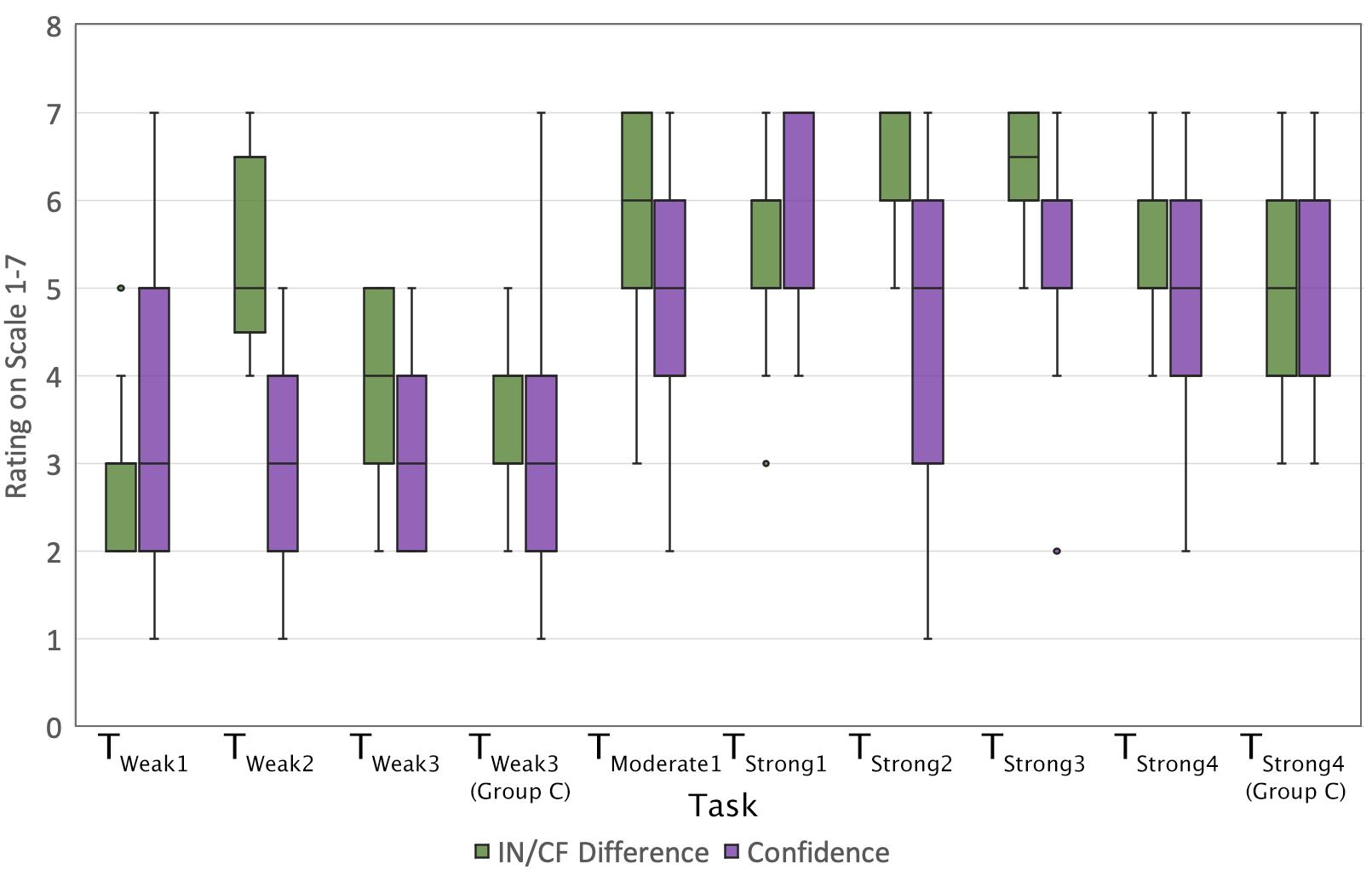}
    \caption{ \label{confidence:diff} Difference between included (IN) and counterfactual (CF) subset outcome distributions and corresponding confidence levels.}
    \vspace{-.5cm}
\end{figure}

\definecolor{olivedrab(web)(olivedrab3)}{rgb}{0.42, 0.56, 0.14}
\definecolor{purplemountainmajesty}{rgb}{0.59, 0.47, 0.71}
\definecolor{napiergreen}{rgb}{0.16, 0.5, 0.0}
\definecolor{ferngreen}{rgb}{0.31, 0.47, 0.26}

Figure~\ref{confidence:diff} displays box plots of the ratings participants provided for the differences in outcome distributions~\tikz\draw[olivedrab(web)(olivedrab3),fill=olivedrab(web)(olivedrab3)] (0,0) circle (.5ex); between the included (IN) and counterfactual (CF) subsets. It shows the corresponding confidence~\tikz\draw[purplemountainmajesty,fill=purplemountainmajesty] (0,0) circle (.5ex); participants reported in the filters' influence on the outcome. 
The 4th and 10th pair of bars (\bm{$T_{\bm{Weak3}}$} and \bm{$T_{\bm{Strong4}}$}, respectively) display group C's responses after viewing the counterfactual visualizations. Generally, if participants reported a lower IN/CF difference rating,
their confidence was
lower as well. 
The exception is \bm{$T_{\bm{Weak2}}$}, for which participants hesitated to report high confidence without knowing how the two applied filters ($F_{SIP}$ + $F_{RNEB}$) interacted, despite reportedly observing a notable difference in the subsets' outcome distributions.

\subsection{User Experience}

\begin{figure}
    \centering
    \includegraphics[height=5.6cm]{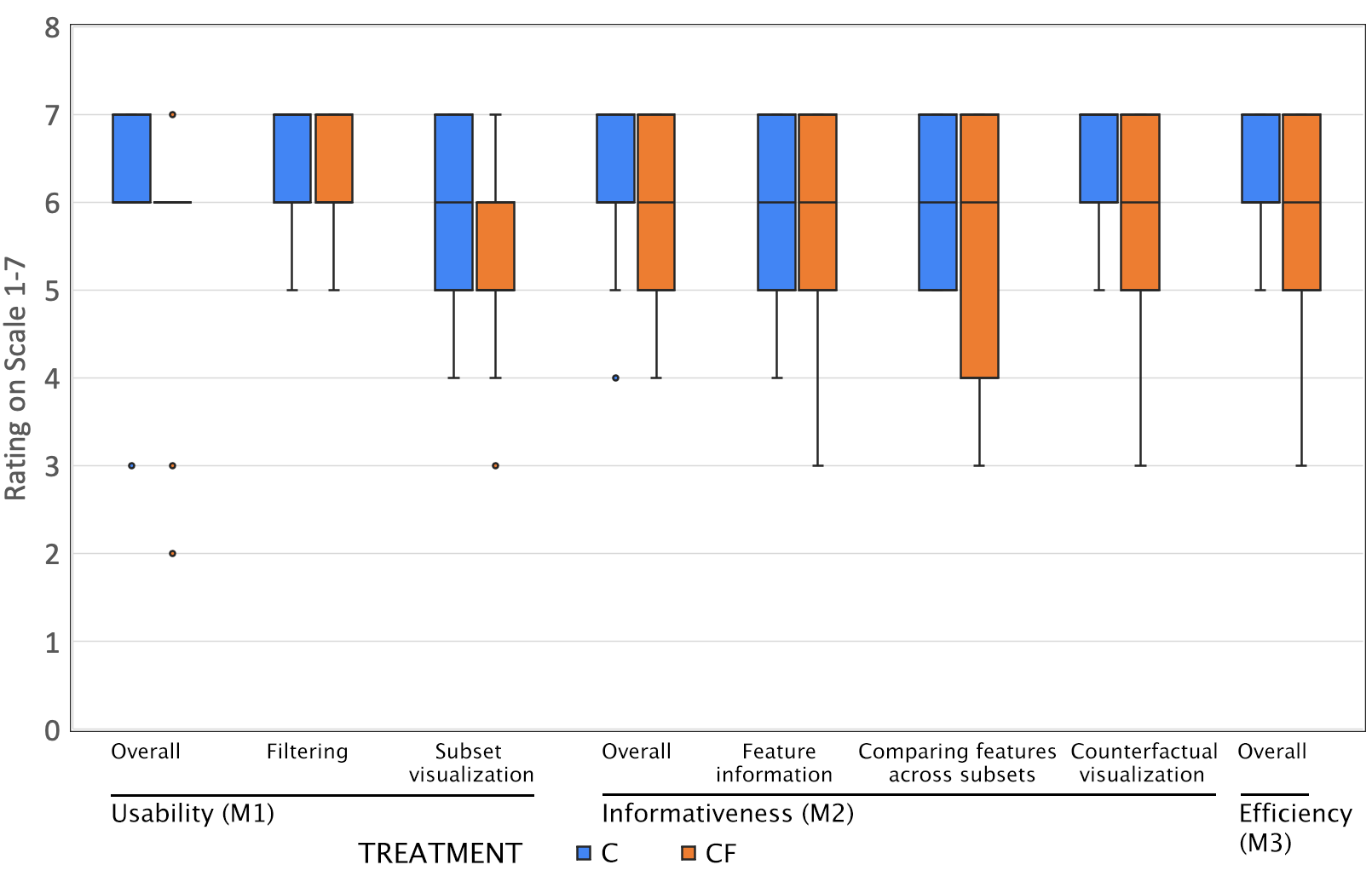}
    \vspace{-.09cm}
    \caption{ \label{feedback} User experience feedback provided by the control (C) and counterfactual (CF) groups for criteria listed in Table~\ref{questions:post}.}
    \vspace{-.5cm}
\end{figure}

After completing the data analysis tasks, participants provided feedback about their experience using CoFact. Each participant rated the extent to which they agreed with the statements related to usability~(\textbf{M1}), informativeness~(\textbf{M2}), and efficiency~(\textbf{M3}) detailed in Table~\ref{questions:post}. Figure~\ref{feedback} displays these responses by groups C~\tikz\draw[azure(colorwheel),fill=azure(colorwheel)] (0,0) circle (.5ex); and CF~\tikz\draw[orange(colorwheel),fill=orange(colorwheel)] (0,0) circle (.5ex);. Overall, ratings were favorable, with median ratings between 5 and 7, where 5 is ``somewhat agree'' that a metric quality (e.g., usability of filtering) was met and 7 is ``strongly agree.'' Across all questions, no significant differences in rating were found between groups C and CF, supporting \textbf{Hypothesis~3}.
In addition to these numeric ratings, the semi-structured post-study interviews captured a wide range of free-form feedback. \hl{Below we summarize findings from a qualitative thematic analysis \cite{braun_using_2006} of the interview responses.}

\textbf{Overall workflow.} 
Participants mostly conveyed positive feedback about the system's overall usability and presentation. 22 of 30 participants (13 in group C, 9 in group CF) specifically noted that the interface was straightforward and easy to use (``gave immediate visual feedback,'' ``very intuitive and easy to use,'' ``user-friendly'') (\textbf{R2}). 12 participants noted that they appreciated the ease of adding and removing filters (\textbf{R1}). They also found useful the correlation information (``that part would be really useful,'' ``ordering by correlation -- I love that'') (\textbf{R1, R3}). 28 participants answered ``Yes'' when asked whether or not the system helped them complete the data analysis tasks, and comments included ``Yes absolutely strongly'' (\textbf{R2}). One participant in group CF who indicated a ``No'' explained that there was a ``lot of confusion.'' 

\textbf{Counterfactual visualization and interpretative assistance.} Overall, participants found the counterfactual visualization valuable (\textbf{R2}). Comments included ``prevents you from jumping to conclusions,'' ``interesting way to look at data that I haven't seen,''
and ``game changer feature.'' The visualization also ``introduced new questions in a good way,'' ``gives more nuance,'' and seemed like ``the voice of reason.'' Group CF participants did express that more effort was required to understand the counterfactual subset (``I found myself getting a bit confused,'' ``I had a little bit of trouble''). Several (11 of 30) asked for additional annotations and interpretation-related indicators. One said it would help if the tool could ``automatically give some hints'' about what the subsets mean.
Another asked for ``confidence intervals, particularly for the counterfactual bar graphs.'' Lastly, participants expressed concern and curiosity about how the counterfactual subset is determined, as well as that they would like to choose what features are used to determine subset similarity: ``more transparency on what's considered similar features,'' ``flexibility on how to calculate that.'' Some suggested that the counterfactual subset could be an optional visualization that users can turn on and off.

\textbf{Customization, additional features, and potential applications.} Three participants from different domain backgrounds noted that they were more familiar with other types of graphs and suggested the ability to choose how data is plotted, i.e., using a pie chart instead of a histogram. 
Five mentioned the ability to customize colors. Also, while most appreciated the ability to choose filter ranges by clicking and dragging on the distribution plots, 11 wanted more precise control and the ability to type values. Two participants noted that CoFact would be a valuable teaching tool to demonstrate certain data analysis and statistical concepts. One said, ``something like this would have a huge impact on an intro [computer science] course,'' ``shows...how you can make more meaning from huge amounts of data.''

\section{Discussion} \label{disc}
This section discusses both key implications and limitations of the user study and CoFact (Section \ref{disc:results}).  Potential future directions for this work are also presented (Section \ref{future}).

\subsection{User Study Results and Limitations} \label{disc:results}

The user study sought to evaluate the three hypotheses described in Section~\ref{hypotheses}, as well as to gather general feedback about user experience with CoFact. Overall, results (Section~\ref{results}) support \textbf{Hypotheses~1--3}: the counterfactual subset visualizations improved user judgment without hampering usability and the overall quality of user experience. Feedback from participant interviews also gave us a more nuanced understanding of this work. Generally, the visualizations helped support the primary requirements \textbf{R1--R3} (Section~\ref{req}): participants indicated that feature information and visualizations helped them choose, apply, and refine filters, after which they were able to, in general, understand the visual subset representations and form judgments about feature-to-outcome relationships. In the future we would allow more time for unguided exploration of the user interface, specifically to more rigorously test \textbf{R1} and \textbf{R3}. 

Two main limitations of the user study were (a)~its time constraint and (b)~the ambiguity of confidence evaluation. First, users had a limited amount of time to be introduced to, digest, and practice the counterfactual approach using the visual system. This likely exacerbated confusion and risked encouraging participants to give quick, non-thorough answers that were truly lower in confidence than was conveyed. Second, participants sometimes found the confidence question (\textbf{Q3} in Table~\ref{questions}) ambiguous. Although the term captures what we intended to study, its ambiguity could have affected participants' responses, and we may seek more precise measures in the future.

\subsection{Future Work} \label{future}

This work used a simple methodology to calculate the counterfactual subset as described in Section~\ref{def}. 
Future work should find more sophisticated methods for this task. In addition, future iterations of this prototype could let users customize various aspects of the counterfactual subset calculation, such as size and alternative similarity metrics, e.g., entropy-based measures. Additionally, several participants asked for greater interpretative support in CoFact. Specifically, participants would like the system to make suggestions about the kind of judgments the user study asked them to make. We would like to explore automating the calculation and communication of visual suggestions about a feature's importance.

\section{Conclusion}
This paper presented a novel counterfactual approach that reveals the presence of confounding factors during visual analysis, implemented in the CoFact prototype system.
CoFact enables users to interactively explore data, apply filter constraints, and analyze the resulting included, excluded, and counterfactual subsets. Visualization of the proposed counterfactual subset, alongside other descriptive information, encourages users to think more critically about feature-to-outcome relationships and the potential of counterfactual possibilities during data exploration.

A controlled user study (n~$=$~30) was conducted to evaluate CoFact,
followed by semi-structured interviews about participants' overall experience. Results indicate that the counterfactual visualizations led to improved
user inference without complicating the interface significantly. With the counterfactual visualizations, users exhibited greater confidence in strong outcome indicators and lower confidence for weak outcome indicators. A thematic analysis of interviews suggested that participants appreciated the counterfactual approach and would find it useful
for data exploration and decision-making. Key areas for future work include (1) greater sophistication and customization in determining counterfactual subsets and (2) calculation and communication of interpretive, actionable counterfactual insights.

\acknowledgments{
The research reported in this article was supported in part by a grant from the National Science Foundation (\#1704018). We also thank Tabitha Peck for her help with data analysis.}

\bibliographystyle{abbrv-doi}

\bibliography{main}
\end{document}